# Vacuumless cosmic strings in Einstein – Cartan theory

[1,3]F. Rahaman, [2]B.C.Bhui, [1]A Ghosh and [1]R. Mondal

[1]Department of Mathematics, Jadavpur University,
Kolkata-700032, India
[2] Dept. of Maths., Meghnad Saha Institute of Technology,
Kolkata-700150, India

[3]E-mail: farook_rahaman@yahoo.com

Abstract:
   The gravitational fields of vacuumless global and gauge strings have been investigated in the context of Einstein – Cartan theory under the weak field assumption of the field equations. It has been shown that global string and gauge string can have only repulsive gravitational effect on a test particle.

Introduction:
   Many interesting developments in cosmology over the last 30 years have come from the realization, due to Kirzhnits [1], that ( almost ) each spontaneous symmetry breaking in Particle physics corresponds to a phase transition in the early Universe. Just like phase transitions in more familiar solids and liquids, cosmological phase transitions can give rise to defects of various kinds. Depending on the topology of the symmetry groups involved, the defects can be in the form of surfaces, lines or points. They are called domain walls, strings and monopoles respectively [2]. The appearance of these structures has produced a great deal of interest because of the cosmological as well as astrophysical implications [3]. In particular, cosmic strings are produced in the breaking of U(1) symmetry, are good candidates to seed the formation of galaxies.
A typical symmetry breaking model is described by the Lagrangian

$$L = \tfrac{1}{2} \partial_\mu \Phi^a \partial^\mu \Phi^a - V(f) \qquad \ldots(1)$$

Where $\Phi^a$ is a set of scalar fields, a = 1, 2,….., N; $f = (\Phi^a \Phi^a)^{1/2}$ and V(f) has a minimum at a non zero value of f . The model has 0(N) symmetry and admits domain wall, string and monopole solutions for N = 1, 2, and 3 respectively. To study the structure of gauge defects, one has to add gauge field in the above Lagrangian and should replace $\partial_\mu$ by a gauge covariant derivatives.

It has been recently suggested by Cho and Vilenkin(CV)[4,5] that topological defects can also be formed in the models where V(f) is maximum at f = 0 and it decreases monotonically to zero for $f \to \infty$ without having any minima.

For example,

$$V(f) = \lambda M^{4+n} (M^n + f^n)^{-1} \qquad \ldots\ldots(2)$$

Where M, $\lambda$ and n are positive constants.
This type of potential can arise in non-perturbative super string models. Defects arising in these models are termed as vacuumless. CV have studied the gravitational field of topological defects in the above models within the frame work of general relativity [5]. In recent, A.A.Sen have studied Vacuumless cosmic strings in Brans-Dicke theory [6]. We know that in general relativity, matter is represented by the energy momentum tensor, which essentially provides a description of mass density distribution in space time. But in particle physics, we know that the matter is formed by elementary particles, which follows the laws of special relativity and quantum mechanics and each particle is characterized not only by a mass, but also by a spin (intrinsic angular momentum). So in general, the matter is described by mass and spin density. In general relativity as given by Einstein there is no way of considering the spin effects on the geometry of space - time. The extension of the geometric principles of general relativity to the physics at a microscopic level where matter formation is done by elementary particles, characterized by a spin angular momentum in addition to mass, is achieved in Einstein – Cartan theory. In a recent paper, we have studied vacuumless global monopole in Einstein – Cartan theory [7]. In this paper, we have studied the gravitational fields field of vacuumless global and gauge strings in Einstein – Cartan theory under weak field approximation of the field equations.

## 2. The Basic equations:

According to CV [4,5] a vacuumless cosmic string is described by a scalar doublet ($\Phi_1$, $\Phi_2$) with a power law potential (2). In cylindrical coordinates ( r,θ,z) , one can assume the ansatz as $\Phi_1 = f(r)\cos\theta$, $\Phi_2 = f(r)\sin\theta$.
For vacuum less global string the flat space-time solution for f(r) is given by [4,5]

$$f(r) = aM(r/\delta)^{2/(n+2)} \qquad \ldots\ldots(3)$$

where $\delta = \lambda^{-1/2} M^{-1}$ is the core radius of the string; r is the distance from the string axis

and

$a = (n+2)^{2/(n+2)} (n+4)^{-1/(n+2)}$. The solution (3) applies for

$$\delta \ll r \ll R \qquad \ldots\ldots(4)$$

where, R is the cut off radius determined by the nearest string.

For gauge vacuumless strings, which have magnetic flux localized within a thin tube inside the core, the scalar field outside the core is given by [4,5]

$f(r) = a \ln (r/\delta) + b$ .........(5)

But here a, b are sensitive to cutoff distance R: $a \sim M(R/\delta)^{2/(n+2)} [\ln (R/\delta)]^{-(n+1)/(n+2)}$

$b \sim a \ln (R/\delta)$.

For a vacuumless cosmic string the space-time is static, cylindrically symmetric and also has a symmetry with respect to Lorenz boost along the string axis. One can write the corresponding line element as

$$ds^2 = A(r)(-dt^2 + dz^2) + A(r) dr^2 + r^2 B(r) d\theta^2 \quad \ldots\ldots(6)$$

The general energy momentum tensor for the vacuumless string is given by

$$T_t^t = T_z^z = \tfrac{1}{2}[(f^1)^2/A] + \tfrac{1}{2}[f^2(1-\alpha)^2/Br^2] + \tfrac{1}{2}[(\alpha^1)^2/Br^2] + V(f) \quad \ldots(7)$$

$$T_r^r = -\tfrac{1}{2}[(f^1)^2/A] + \tfrac{1}{2}[f^2(1-\alpha)^2/Br^2] - \tfrac{1}{2}[(\alpha^1)^2/2Br^2] + V(f) \quad \ldots\ldots(8)$$

$$T_\theta^\theta = \tfrac{1}{2}[(f^1)^2/A] - \tfrac{1}{2}f^2[(1-\alpha)^2/Br^2] - \tfrac{1}{2}[(\alpha^1)^2/Br^2] + V(f) \quad \ldots\ldots(9)$$

The string ansatz for the gauge field is $A_\theta(r) = -\alpha(r)/er$

where *e* is the strength of the gauge coupling.

$T_a^b$'s with $\alpha = 0$ are that for global string.

Following Prasanna [8], the Einstein – Cartan equations can be written as

$$R_a^b - \tfrac{1}{2} R \delta_a^b = -8\pi G T_a^b \quad \ldots\ldots(10)$$

$$Q_{bc}^a - \delta_b^a Q_{lc}^l - \delta_c^a Q_{bl}^l = -8\pi G S_{bc}^a \quad \ldots(11)$$

Here we have assumed that the spins of the individual particles are aligned along the symmetry axis (z axis).
So only nonzero components of the spin tensor $S_{ij}$ is $S_{r\theta} = -S_{\theta r} = K(\text{say})$ ..(12)
The non zero components of $S_{bc}^a$ are

$S_{r\theta}^t = -S_{\theta r}^t = K$

....(13)

[ here, $S_{bc}^a$ is the spin density described through the relation $S_{bc}^a = U^a S_{bc}$
with $U^c S_{bc} = 0$, where $U^a$ is the four velocity vector $U^a = \delta_t^a$ ]

Consequently, from Cartan equations (11), we get the torsion tensor $Q_{bc}{}^a$ to be

$$Q_{r\theta}{}^t = - Q_{\theta r}{}^t = - 8\pi GK \qquad \ldots\ldots(14)$$

where the other components are zero.

The filed equations for the metric (6) in Einstein – Cartan theory are

$$\tfrac{1}{2}(A^1 B^1 / A^2 B) + \tfrac{1}{4}[(A^1)^2/A^3] + (A^1/rA^2) + (1/A)16\pi^2 G^2 K^2 = 8\pi G\, T_r^r \qquad \ldots(15)$$

$$-\tfrac{3}{4}[(A^1)^2/A^3] + (A^{11}/A^2) + (1/A)16\pi^2 G^2 K^2 = 8\pi G\, T_\theta^\theta \qquad \ldots(16)$$

$$-\tfrac{1}{2}[(A^1)^2/A^3] + \tfrac{1}{2}(A^{11}/A^2) + \tfrac{1}{2}(B^{11}/B^2 A) + (B^1/rAB) - \tfrac{3}{4}[(B^1)^2/AB^2]$$

$$- (1/A)16\pi^2 G^2 K^2 = 8\pi G\, T_t^t \qquad \ldots..(17)$$

[ '$^1$' indicates differentiation w.r.t. r ]

## 3. Global String:

Under the weak field approximations one can write

$$A(r) = 1 + \beta(r) \text{ and } B(r) = 1 + \gamma(r) \qquad \ldots..(18)$$

Where $\beta, \gamma \ll 1$. For global vacuumless string, one can use the flat space approximation for f(r) in (3) for $r \gg \delta$ and the form of V(f) given in (2).
Now under these weak field approximations, the field equations take the following forms:

$$(\beta^1/r) + 16\pi^2 G^2 K^2 = D[\,(n^2 + 6n + 16)/(n+2)^2\,]\, r^{-b} \qquad \ldots..(19)$$

$$\beta^{11} + 16\pi^2 G^2 K^2 = D[\,(n-4)/(n+2)\,]\, r^{-b} \qquad \ldots..(20)$$

$$\tfrac{1}{2}\beta^{11} + \tfrac{1}{2}\gamma^{11} + (\gamma^1/r) - 16\pi^2 G^2 K^2 = D[\,(n+4)/(n+2)\,]\, r^{-b} \qquad \ldots..(21)$$

where $D = 8\pi G\, a^2 M^2 \delta^{-4/(n+2)}$ and $b = 2n/(n+2)$.

From eq.(19), we get the following solution of $\beta$ as

$$\beta = -8\pi^2 G^2 K^2 r^2 - D[\,(n^2 + 6n + 16)/(2-b)(n+2)^2\,]\, r^{-b+2} \qquad \ldots\ldots(22)$$

Also from eq.(20), we get the following solution of β as

$$\beta = - 8\pi^2 G^2 K^2 r^2 - D[(n-4)/(1-b)(2-b)(n+2)] r^{-b+2} \qquad \ldots\ldots(23)$$

For consistency, we must have that the second term of both the equations is same i.e.
n is connected with the consistency relation $n^3 + 2n^2 - 4n - 24 = 0$.
One can see that n lies between 2 and 3.
For n > 2, the finite temperature potential always has a local minimum at f = 0.
However, this minimum is protected by a potential barrier whose width and height rapidly decrease with the temperature. This barrier can be easily overcome by thermal fluctuations [9].
Also we get the solution of γ as

$$\gamma = 8\pi^2 G^2 K^2 r^2 + D[(n+12)/(3-b)(2-b)(n+2)] r^{-b+2} \qquad \ldots\ldots(24)$$

## 4. Gauge String:

For gauge vacuumless string the energy momentum tensor with f(r) given in (5) can be approximated as [4,5]

$$T_t^t = T_z^z = T_\theta^\theta = -T_r^r = \tfrac{1}{2}(f^1)^2 = \tfrac{1}{2}[a^2/r^2] \qquad \ldots..(25)$$

This form of the energy momentum is valid for

$$r << R[\ln(R/\delta)]^{-\tfrac{1}{2}} \qquad \ldots\ldots.(26)$$

where R is the cutoff radius determined by the nearest string.

For gauge string using the weak field approximations (18), the field equations read

$$(\beta^1/r) + 16\pi^2 G^2 K^2 = -4\pi G [a^2/r^2] \qquad \ldots..(27)$$

$$\beta^{11} + 16\pi^2 G^2 K^2 = 4\pi G [a^2/r^2] \qquad \ldots..(28)$$

$$\tfrac{1}{2}\beta^{11} + \tfrac{1}{2}\gamma^{11} + (\gamma^1/r) - 16\pi^2 G^2 K^2 = 4\pi G [a^2/r^2] \qquad \ldots..(29)$$

Solving these equations, we get,

$$\beta = -8\pi^2 G^2 K^2 r^2 - 4\pi G a^2 \ln r \qquad \ldots..(30)$$

$$\gamma = 8\pi^2 G^2 K^2 r^2 + 4\pi G a^2 \ln r \qquad \ldots..(31)$$

## 5. Gravitational effects on test particles:

The repulsive and attractive character of the global string can be discussed by either studying the time like geodesics in the space-time or analyzing the acceleration of an observer who is rest relative to the string.

We now calculate the radial acceleration vector $A^r$ of a particle that remains stationary (i.e. $V^1 = V^2 = V^3 = 0$) in the field of the string.

Let us consider an observer with four velocity $V_i = \sqrt{(g_{00})}\delta_i^t$.

[ 0 and 1 stand for t and r respectively ]

Now, $A^r = V^1{}_{;0} V^0 = \Gamma_{00}^1 V^0 V^0$ ....(32)

For global string,

$$A^r = [-16\pi^2 G^2 K^2 r - D\{(-n+4)/(b-1)(n+2)\} r^{-b+1}](1+\beta)^{-3} \quad ....(33)$$

We see that $A^r$ is a function of 'r' and for n lies between 2 and 3, the expression $A^r$ is negative. Thus in this case gravitational force varies with the radial distance and one can have repulsive gravitational effect on the test particle.

For gauge string,

$$A^r = [-16\pi^2 G^2 K^2 r - (4\pi G a^2/r)](1+\beta)^{-3} \quad ....(34)$$

Here, the acceleration vector is always negative and gravitational force is repulsive.

## 6. Conclusions:

In conclusion, this work extends the earlier work by Cho and Vilenkin regarding the gravitational field of vacuumless cosmic strings to Einstein – Cartan theory. We see that in going from general relativity to Einstein – Cartan theory both space time curvature and topology are affected by the presence of spin tensor. Study of the motion of the test particle reveals that the vacuumless global string in Einstein – Cartan theory exerts gravitational force, which is repulsive in nature. It is similar to the case of a vacuumless global string in general relativity where the vacuumless global string has only repulsive gravitational effect [5] but dissimilar to the case studied in Brans-Dicke theory [6].
In case of gauge string, we see that vacuumless gauge string in Einstein – Cartan theory exerts gravitational force, which is only repulsive in nature. The vacuumless gauge string in general relativity exerts attractive gravitational force near the string and repulsive far away [5]. So this observation is in striking contrast with the analogue in Einstein's theory. Recently, Rahaman et al [10] have shown that global string in Einstein-Cartan theory exhibits repulsive gravitational force. Thus it seems the spin in the Einstein-Cartan theory is responsible of these repulsive forces.

Acknowledgements:

We are grateful to Dr.A.A.Sen for helpful discussions. F.R is thankful to IUCAA for providing research facility. We are also grateful to the referees for their constructive suggestions to improve the manuscript.